\documentclass[article,twocolumn]{revtex4}

\usepackage{graphicx}
\usepackage{amsmath}
\usepackage{amssymb}
\usepackage{fancyhdr}
\usepackage{amsfonts}
\usepackage{multirow}
\usepackage{color}

\newcommand{\Tr}{\mathrm{Tr}}

\newcommand{\be}{\begin{equation}}
\newcommand{\ee}{\end{equation}}
\newcommand{\bea}{\begin{eqnarray}}
\newcommand{\eea}{\end{eqnarray}}

\newcommand{\bean}{\begin{eqnarray*}}
\newcommand{\eean}{\end{eqnarray*}}

\newcommand{\expect}[1]{\left\langle#1\right\rangle}

\date{\empty}

\begin{document}
\title{Detecting correlated errors in SPAM tomography}
\author{Christopher Jackson and S.J.~van Enk}
\affiliation{Oregon Center for Optics and
Department of Physics\\
University of Oregon, Eugene, OR 97403}
\date{\today}
\begin{abstract}
Whereas in standard quantum state tomography one estimates an unknown state by performing various measurements with known devices; and whereas in detector tomography one estimates the POVM elements of a measurement device by subjecting to it various known states, we consider here the case of SPAM (state preparation and measurement) tomography where neither the states nor the measurement device are assumed known. For $d$-dimensional systems measured by $d$-outcome detectors, we find there are at most 
$d^2(d^2-1)$ ``gauge'' parameters that can never be determined by any such experiment, irrespective of the number of unknown states and unknown devices. For the case $d=2$ we find new gauge-invariant quantities that can be accessed directly experimentally and that can be used to detect and describe SPAM errors.
In particular, we identify conditions whose violations detect the presence of correlations between SPAM errors. From the perspective of SPAM tomography, standard quantum state tomography and detector tomography are protocols that  fix the gauge parameters through the  assumption that some set of fiducial measurements is known or that some set of fiducial states is known, respectively.
\end{abstract}
\maketitle
\section{Introduction}
Quantum tomography has become an important tool for characterizing quantum devices \cite{raymer,paris}.
For example, in order to estimate the state $\rho$ of quantum systems produced by a given quantum source, we let the source create many copies of $\rho$, and subject  those copies to different measurements.
If we describe the whole set of measurements by POVM elements $\Pi_i$
(such that $\sum_i \Pi_i=\openone$), the experiment will produce estimates of probabilities
\be
p_i=\Tr(\rho\Pi_i),
\ee
with $\sum_i p_i=1$.
If we have at least as many observed independent relative frequencies $f_i$ (the number of times outcome $i$ was observed divided by the total number of measurements) as there are parameters in $\rho$, then we may estimate $\rho$ via a variety of different methods (from linear inversion to hedged maximum likelihood estimation  and Bayesian methods, see, e.g., \cite{blumeh,blumeb,schwemmer}).
Legitimately ending up with an estimate of a {\em single} density matrix $\rho$ does require some assumptions about the experimental setup \cite{finetti,renner,blume}, namely, invariance under permutations of the copies. In particular,  even if there is drift (for example, due to a slowly changing magnetic field or a slowly changing phase of the laser field used to prepare our quantum systems), such that copy $\rho_n$ differs slightly from the previous copy $\rho_{n-1}$, we still obtain an estimate of a single density matrix, namely, the average:
\be\label{rhoav}
\overline{\rho}=\frac{1}{N_t}\sum_{n=1}^{N_t} \rho_n,
\ee 
in an experiment with $N_t$ copies of states in total, provided the different measurements are performed in random order.

In more recent times several experiments on detector tomography have been performed \cite{lundeen,feito,renema,renemab}.
Here the assumption is that known states are fed into an unknown measurement device, whose observed outcomes now tell us about the POVM elements that describe the detector. 
Detector tomography makes sense: A photodetector, for example, is not such a simple device theoretically. There are infinitely many modes of the electromagnetic field, and each mode is described by an infinite-dimensional Hilbert space.
A simple candidate model for a photodetector can be formulated in terms of a quantum efficiency $\eta$ and a dark count rate $r$, but these two parameters could drift over time and be field-mode dependent.
In the case of detector tomography, we  estimate probabilities of the form
\be
p_\mu=\Tr(\rho_\mu\Pi),
\ee
for different known states $\rho_\mu$ (where $\mu$ runs from 1 to $M$), in order to estimate the POVM element $\Pi$, where the latter is really an average (under the proviso that we test the measurement device with different states created in random order, to ensure symmetry under permutations of the measurements) over the course of the experiment of the different POVM elements 
$\Pi_n$ describing the drifting detector. In complete analogy to (\ref{rhoav}) we may write then
\be
\overline{\Pi}=\frac{1}{N_t}\sum_{n=1}^{N_t} \Pi_n,
\ee
if we performed this particular measurement $N_t$ times in total.

The present paper focuses on two questions: First, what exactly can we find out about measurements and states when we do not assume we know either one or the other?
That is, what can we infer about the POVM elements $\Pi_i$ describing different measurement settings and about different states $\rho_\mu$ from estimates of probabilities
\be
p_{\mu,i}=\Tr(\rho_\mu\Pi_i)?
\ee
Second, what changes if the fluctuations (or drift) in state preparation and measurement are correlated? That is, what if the observed average relative frequencies are not determined by averaged states and averaged measurements? How can we infer there are correlations such that
\be
\overline{\Tr(\rho_\mu\Pi_i)}\neq
\Tr(\overline{\rho}_\mu\overline{\Pi}_i)?
\ee
Here the left-hand side is a quantity for which we obtain direct estimates from observed relative frequencies, but the two quantities $\overline{\rho}_\mu$ and $\overline{\Pi}_i$ on the right-hand side are not directly accessible individually and have to be inferred.

Such questions have become of interest in the context of debugging quantum devices that are meant to serve as fault tolerant quantum computers.
The requirements on fault tolerance are quite stringent and more and more precise tools for analyzing tomography experiments have been developed very recently \cite{gst,merkel,stark}. Correlations between errors are particularly bad for fault tolerance, hence our focus on detecting correlated SPAM errors.

We are going to analyze SPAM tomography for the case of a single qubit and two-outcome POVMs. Even for this simple case our results suggest novel experiments. Most results generalize easily to multiple qubits and/or higher-dimensional systems and measurements, and in particular to the combination of $d$ dimensional systems and $d$-outcome POVMs.
\section{SPAM tomography}
\subsection{Notation}
To set the notation, let us first consider the simplest version of our problem: a single qubit state, $\rho$, and a single two-outcome qubit detector, with the two outcomes described by POVM elements $\{E,\neg E\}$ (read: $E$ and NOT-$E$).
From the conditions $\rho\ge 0, E\ge 0, \neg E \ge 0$, $\Tr\rho =1$, and $E+\neg E = \openone$, a general parameterization can be written in terms of the Pauli matrices $\vec\sigma=(\sigma_x,\sigma_y,\sigma_z)^T$ as
\bea
	\rho &=& \frac{1}{2}\big(\openone+\vec p \cdot\vec{\sigma}\big),\nonumber\\
	E &=& \frac{1}{2}\Big((1+ u)\openone+\vec w \cdot \vec{\sigma}\Big),\nonumber\\
\neg E &=& \frac{1}{2}\Big((1- u)\openone-\vec w \cdot \vec{\sigma}\Big),
\eea
where positivity is ensured by the inequalities
\bea
\label{Sigma}
	|\vec p| &\leq& 1,\nonumber\\
	|\vec w| + |u| &\leq& 1.
\eea
Altogether there are seven parameters. Three of these correspond to a choice of coordinate system (when we think of the qubit as a spin-1/2 system, we have to specify what we mean by the components of spin in the $x,y,z$ directions), or, equivalently, to a choice of basis for the 2D Hilbert space describing our qubit.
The other four parameters can be thought of as  state purity ($|\vec p|$), the detector's discrimination power ($|\vec w|$), detector bias ($u$), and the alignment (or fidelity) between detector and state ($\hat p \cdot \hat w$).
We may conveniently represent the two-outcome POVM by a single observable:
\begin{equation}\label{obs}
	\Sigma \equiv E-\neg E = u\openone+\vec w \cdot \vec{\sigma}.
\end{equation}
Knowledge of its expectation value $S$ in the state $\rho$ gives us a single constraint on these seven parameters:
\begin{equation}\label{constraint}
	S \equiv \Tr\rho\Sigma = \vec p \cdot \vec w + u.
\end{equation}

\subsection{Counting parameters}
Now consider the more interesting case
where we perform two-outcome measurements with a variety of states and observables.
We press the $\mu$-th button of a quantum source, where $\mu$ runs from 1 to $M$, representing the preparation  of state $\rho_\mu$, while the dial of the measurement device is turned to the $i$-th setting (where $i$ runs from 1 to $N$), symbolizing the observable $\Sigma_i$. The outcome of measurement $i$ is either that an indicator light blinks ($E_i$) or does not blink ($\neg E_i$).

Considering the number of undeterminable parameters, $\gamma$, a first guess might be that there are
\begin{equation}
\gamma=	(\text{\# unknowns}) - (\text{\# constraints}) = 3M+4N-NM
\end{equation}
of them, where we assume we always subject all $M$ states to all $N$ measurements.
This kind of counting is correct so long as either $N< 3$ or $M<4$ but otherwise it fails to take the following important observation into account:
For $M=4$, there are sufficiently many states to perform complete detector tomography on any number of detectors. Similarly, $N=3$ would be a sufficient number of 2-outcome detectors to perform complete state tomography.
Hence for any $N>3$ and $M>4$, there are just as many undeterminable parameters as when $N=3$ and $M=4$,
because the addition of any states (or detectors) could reveal no more information about the detectors (or states) than the ``original'' 4 states (or 3 detectors) already yield. (This argument generalizes to $d$-outcome measurements performed on $d$-dimensional systems: each $d$-outcome measurement yields $d-1$ independent frequencies. $d+1$ known $d$-outcome POVMs are, therefore, sufficient to determine the $d^2-1$ parameters of any unknown state. 
Conversely, $d^2$ known states (each needing $d^2-1$ parameters for its specification) are needed to determine the $d^2$ parameter of any unknown POVM element. Thus, the maximum number of gauge parameters is $d^2(d^2-1)$.)
Therefore the number of undeterminable parameters is
\begin{equation}
	\gamma =
	\left\{
	\begin{array}{crl}
		3M+4N-NM & : & N<3 \text{ or } M < 4\\
		12 & : & N\geq 3 \text{ and } M \geq 4\\
	\end{array}
	\right.
\end{equation}
(From now on we assume we do have at least 4 states and at least 3 detectors, so that the number of undeterminable parameters is 12.)
In order to find what these undeterminable parameters are we first introduce some convenient notation.
First of all, the set of all constraints can be written succinctly as a matrix equation,
\begin{equation}\label{simple}
S = 
PW
\end{equation}
where
\be\label{lists}
P
\equiv
\left[
\begin{array}{ccc}
\cdots & \vec p_\mu & \cdots \\
	 \cdots& 1 &  \cdots
\end{array}
\right]^{T}
\ee
and
\be
W
\equiv
\left[
\begin{array}{ccc}
	\cdots & \vec w_i & \cdots \\
	\cdots & u_i & \cdots
\end{array}
\right]
\ee
are the $N \times 4$ and $4 \times M$ matrices of state and detector parameters, respectively, where we included a constant 1 (which stands for $\Tr(\rho_\mu)$) in the set of state parameters just to make the treatment of states and detectors more symmetric.
The form of equation (\ref{simple}) makes obvious that the system of constraints has the symmetry:
\begin{equation}
	P \longrightarrow PG^{-1},
	\,\,\,\,
	W \longrightarrow GW,
\end{equation}
where $G$ is of the general form
\begin{equation}\label{GH}
G=
\left[
\begin{array}{cc}
	H & \vec 0 \\
	\vec a^T & 1
\end{array}
\right]
\end{equation}
where $H$ is any real $3\times3$ invertible matrix, $H \in {\rm GL}(3)$, and $\vec a$ is any real 3-vector, $\vec a \in \mathbb{R}^3$.
The collection of all matrices $G \in \mathbb{R}^3 \rtimes GL(3)$ forms the affine group.

\subsection{Blame Gauges: A New Perspective on State/Detector Tomography}
Some of the 12 parameters determining $G$ should be familiar.
Most important are perhaps the three parameters needed to describe matrices of the form
\begin{equation}\label{R}
G=
\left[
\begin{array}{cc}
	R & \mathbf0 \\
	\mathbf0 & 1
\end{array}
\right].
\end{equation}
where $R$ is in the adjoint representation of ${\rm SU}(2)$ (which here is just ${\rm SO}(3)$.)
This space of symmetries is always present (also for higher-dimensional Hilbert spaces of size $d$ where we would have the adjoint representation of SU($d$)), and corresponds to the arbitrary choice of basis for the underlying Hilbert space. This sort of symmetry tends to be trivial, but becomes nontrivial (and interesting)
in the context of quantum communication
(the information to fix a reference frame was coined ``unspeakable" by Asher Peres \cite{peres}, for the reason that two distant parties cannot communicate the choice of coordinate system over the phone, unless they already share a reference frame).

Another simple undeterminable parameter is revealed by considering $G$ of the form 
\begin{equation}\label{scale}
G=
\left[
\begin{array}{cc}
	g\openone_3 & \mathbf0 \\
	\mathbf0 & 1
\end{array}
\right],
\end{equation}
where $g \in {\mathbb R}$.
This kind of scaling transformation trades overall state purity for the detector's discrimination power. That is, imperfections may be blamed, within some limits, more on the preparation of the states than on the quality of our detectors or {\em vice versa} \footnote{This parameter is sometimes referred to as the SPAM gauge.}.
Let us refer to parameters like this as {\em blame gauge} degrees of freedom.

We can characterize all 12 blame gauge parameters as follows.
Every $H$ (as appearing in (\ref{GH})) has a polar decomposition $H= B R$ where $R$ is real orthogonal and $B$ is positive symmetric.
This yields the following interpretation:
We can always diagonalize $B$: the three eigenvalues are like the scale parameter $g$ of (\ref{scale}). The basis in which $B$ is diagonal (determined by three additional parameters)
determine which components of $\vec{p}$ and $\vec{w}$ can be rescaled in concert.

The three parameters of $R$ still correspond to the choice of Hilbert space basis (or of a coordinate system) as in (\ref{R}) \footnote{The interpretation of $B$ and $R$ is more subtle for higher Hilbert space dimensions, $d>2$, because we loose the ``accidental isomorphism'' $SO(3)\cong SU(2)/Z_2$.}.

Finally, the last three parameters  feature in transformations of the form
\begin{equation}
G=
\left[
\begin{array}{cc}
	\openone_3 & \mathbf0 \\
	\vec a^T & 1
\end{array}
\right],
\end{equation}
where $\vec a$ is a real 3-vector.
It is perhaps clearer what this transformation does when we write out the action on the system parameters explicitly:
\begin{equation}
	\left\{
	\begin{array}{l}
		\vec p_\mu \longrightarrow \vec p_\mu-\vec a\\
		\vec w_i \longrightarrow \vec w_i\\
		u_i \longrightarrow u_i+\vec a\cdot\vec w_i.
	\end{array}
	\right.
\end{equation}
We see that this exchanges state-detector alignment for detector bias, and thus is another type of blame gauge.

Having now completely classified the 12 gauge degrees of freedom in qubit SPAM tomography, let us briefly revisit the more conventional types of (qubit) tomography.
Having at our disposal $M=4$ fiducial states assumed to be fully known allows for complete detector tomography precisely because we can write in that case $W=P^{-1}S$.
Similarly, $N=3$ known fiducial detectors allows us to perform complete quantum state tomography because we can write in that case $P = SW^{-1}$, where $W^{-1}$ is to be interpreted as a right (pseudo-)inverse.

From the perspective of SPAM tomography we see, therefore, that the assumptions underlying state and detector tomography  simply boil down to fixing the {\em blame gauge}.
In state tomography on qubits one fixes four parameters each of three fiducial measurements, and in detector tomography one fixes four parameters each of three fiducial states.
\section{Detecting Correlated Errors}
\subsection{Uncorrelated States and Detectors}
A measurement yields averages of the form
\begin{equation}\label{ontolo}
	S_{\mu i} =\overline{\expect{\Sigma_i}}_\mu=\overline{\vec p_\mu \cdot \vec w_i + u_i}, 
	\end{equation}
	which contains two types of averages: one quantum expectation value (in the state $\rho_\mu$, denoted as usual by $\expect{.}$), and one average over the different runs of the experiment (indicated by the over line).
The existence of a model $(P,W)$ that satisfies Eq.~(\ref{simple}) is precisely the statement that a system consists of \emph{effectively uncorrelated} states and detectors.
The point is now that, certainly for $M > 4$ and $N > 3$, not every possible set of data can be effectively uncorrelated simply because
the number of data parameters is larger than the number of distinct uncorrelated model parameters whenever
\begin{equation}
 MN > 3M+4N-12. 
\end{equation} 
We will now answer the question under what conditions there is an uncorrelated model and when there is not.


\subsection{Constraints for Uncorrelated Quantum Data: Partial Determinants}

Suppose we have quantum data $S$ that is effectively uncorrelated, i.e., described by a model $(P,W)$.
For the sake of presentation, let us consider first the case $N=M=8$.
Think of the data as partitioned into four $4 \times 4$ corners of an $8\times 8$ matrix,
\begin{equation}\label{corners}
S
\equiv
\left[
\begin{array}{cc}
	Q_{11} & Q_{12}\\
	Q_{21} & Q_{22}
\end{array}
\right],
\end{equation}
such that
\begin{equation}\label{persp}
Q_{ab}=P_aW_b
\end{equation}
where
\begin{equation}
P
=
\left[
\begin{array}{c}
	P_1 \\ P_2
\end{array}
\right];\,\,\,
W =
\left[
\begin{array}{cc}
	W_1 & W_2
\end{array}
\right],
\end{equation}
| that is, 
\begin{equation}
P_1 = 
\left[
\begin{array}{ccc}
	\vec p_1 &\cdots & \vec p_4\\
	1 &\cdots  & 1
\end{array}
\right]^T;\,\,\,\,
P_2 = 
\left[
\begin{array}{ccc}
	\vec p_5 & \cdots & \vec p_8\\
	1 &  \cdots & 1
\end{array}
\right]^T
\end{equation}
are the top and bottom halves of $P$ and 
\begin{equation}
W_1 =
\left[
\begin{array}{ccc}
	\vec w_1 & \cdots& \vec w_4\\
	u_1 & \cdots & u_4
\end{array}
\right];\,\,\,\,
W_2 =
\left[
\begin{array}{ccc}
	\vec w_5 & \cdots & \vec w_8\\
	u_5 & \cdots & u_8
\end{array}
\right]
\end{equation}
are the left and right halves of $W$.

These submatrices of the data, because of Eq.~(\ref{persp}), have the property
\begin{equation}\label{pds}
	Q_{11}^{-1} Q_{12} Q_{22}^{-1}  Q_{21} =\openone_4,
\end{equation}
under the assumption that the two inverse matrices on the left-hand side exist.
(For example, when state 1 and state 2 are  the same, or measurements 1 and 3 are  the same,  $Q_{11}^{-1}$ does not exist. Also,  if  one of the four vectors $\vec{p}_i$ for $i=1,2,3,4$ is a mixture of the other three, then $Q_{11}^{-1}$ does not exist either.)
We will refer to the left hand side of (\ref{pds}) as a {\em partial determinant} of $S$ and write
\begin{equation}\label{pd}
	\Delta(S) \equiv Q_{11}^{-1}Q_{12} Q_{22}^{-1} Q_{21}.
\end{equation}
One should note that the ordering of the rows of $P$ and columns of $W$ was arbitrary and any such reordering would result in a different partial determinant.
Note that a partial determinant can be defined for values of $N$ and $M$ as low as $5$ \footnote{
It is worth mentioning that the statement $\Delta(S) = 1$ for all combinations of measurements and states is generically equivalent to the statement that $\mathrm{rank}(S) \le 4$.
Geometrically, partial determinants are interesting because they formalize the subspace of effectively uncorrelated data as a variety | simply, solutions to the system of equations $\Delta = 1$ (over various $\Delta$s from different permutations of the data.)
}.

The important conclusion is that the condition that $\Delta{S}=\openone_4$
can be used to detect correlations: a violation of this equality (while taking into account error bars) shows that state preparation and measurement must have been correlated.
\subsection{Nonsingular Partial Determinants}
To simplify notation, instead of using $Q_{ab}$s, let us write the partitions of the data, $S$, simply as
\begin{equation}
S
\equiv
\left[
\begin{array}{cc}
	A & B\\
	C & D
\end{array}
\right]
\end{equation}
in which case we have
\begin{equation}
	\Delta(S) = A^{-1} B D^{-1}  C.
\end{equation}
An awkward feature of this quantity is that it  may display singular behavior because of the presence of inverse matrices.
A simple way to remedy this is to define a {\em nonsingular partial determinant}
\footnote{This quantity is in fact proportional to what's called the Schur complement, $A-B D^{-1} C$, of $A$.}
\begin{equation}\label{nspd}
	\nabla(S) \equiv \det(D) A\big(\Delta - \boldsymbol1\big) = B \bar D^T C - (\det D)A
\end{equation}
where $\bar D$ is the cofactor matrix of D so that, for example,
\begin{equation}
	A^{-1} = \frac{1}{\det A}\bar A^T.
\end{equation}
We chose to multiply by $A$ simply to reduce computational complexity.
This then gives rise to another (non-singular) condition on having uncorrelated SPAM errors.
It has the disadvantage that satisfaction of the condition $\nabla(S)=0$ may occur not because one's SPAM errors are uncorrelated, but because of having multiplied both sides of (\ref{pds}) with zero.
\subsection{Estimates and Uncertainty for the Partial Determinant of Qubit Data}\label{est}
We consider here briefly error bars in the partial determinant, which are needed to decide whether or not one's data provides significant evidence for correlated SPAM errors.

The data is estimated to maximize the likelihood of $MN$ binomial distributions:
\begin{equation}
	\langle S\rangle_{\mu i} \equiv \langle S_{\mu i} \rangle = 2 \frac{k_{\mu i}}{n_{\mu i}}-1
\end{equation}
where $n_{\mu i}$ is the number of trials for the $\mu$-th state and $i$-th detector settings and $k_{\mu i}$ is the number of positive outcomes.
As such, the deviation of values from the actual mean is
\begin{equation}
	(\delta S_{\mu i})^2 \equiv \langle {S_{\mu i}}^2\rangle - \langle S_{\mu i}\rangle^2  = \frac{1-{S_{\mu i}}^2}{n_{\mu i}}.
\end{equation}
We can then propagate these uncertainties to the nonsingular partial determinant,
\begin{equation}\label{nspd}
	\nabla(S) = B \bar D^T C - (\det D)A.
\end{equation}
Assuming independent Gaussian random variables, the relative coefficients for the qubit case are
\begin{equation}
	\frac{\partial \nabla_{ab}}{\partial A_{cd}} = -\delta_{ac}\delta_{bd}\det D
\end{equation}
\begin{equation}
	\frac{\partial \nabla_{ab}}{\partial B_{cd}} = \delta_{ac}\varepsilon_{ijk}\varepsilon_{dlm}C_{ib}D_{jl}D_{km}
\end{equation}
\begin{equation}
	\frac{\partial \nabla_{ab}}{\partial C_{cd}} = \delta_{bd}\varepsilon_{cij}\varepsilon_{klm}B_{ak}D_{il}D_{jm}
\end{equation}
\begin{equation}
	\frac{\partial \nabla_{ab}}{\partial D_{cd}} = \varepsilon_{ijc}\varepsilon_{kld}\left(B_{ak}D_{jl}C_{ib}-\frac{1}{2}D_{ik}D_{jl}A_{ab}\right)
\end{equation}
where $\delta$ and $\varepsilon$ are the Kronecker and Levi-Civita symbols, reps., and repeated indices are to be summed over.
If we choose $N=M=8$, then the subdata can be chosen to be statistically independent and the error bar in the partial determinant is then simply
\bea
	(\delta\nabla_{ab})^2 = \sum_{c,d}\left(\frac{\partial \nabla_{ab}}{\partial A_{cd}} \delta A_{cd}\right)^2+\left(\frac{\partial \nabla_{ab}}{\partial B_{cd}} \delta B_{cd}\right)^2\nonumber\\
		+\left(\frac{\partial \nabla_{ab}}{\partial C_{cd}} \delta C_{cd}\right)^2+\left(\frac{\partial \nabla_{ab}}{\partial D_{cd}} \delta D_{cd}\right)^2.
\eea
Otherwise, if a smaller number of states and detectors is considered, one must take into account that the entries are not all statistically independent when calculating error bars.

\subsection{Examples of Correlated Models}

A significant feature of these partial determinants is that they can be estimated directly from measured data.
Stressing the point explicitly:
``$\nabla(S) = 0$'' may serve as an operational definition for $S$ to have no state-detector correlations.
Certainly, if one finds that $\nabla(S) \neq 0$ with high confidence, then one has made a definitive statement that there are state-detector correlations.

It is perhaps helpful if we illustrate the concept of correlated errors with a couple of simple models.
For the sake of simplicity, let us restrict our attention to measurement devices whose  biases $u_i$ may be chosen to all vanish.
\subsubsection{Gaussian model}
A first way to model correlation is to treat the Bloch vectors for all states and detectors as randomly fluctuating:
\begin{equation}
\begin{array}{c}
	\vec{p}_\mu = \vec{p}_{\mu o} + \vec f_\mu\\
	\vec{w}_ i = \vec{w}_{i o} + \vec g_i\\
\end{array}
\end{equation}
where the $\vec{p}_{\mu o}$ and $\vec{w}_{i o}$ are constants and $\vec f_\mu$ and $\vec g_i$ are Gaussian noise with
\begin{equation}
	\langle\vec f_\mu\rangle = \langle\vec g_i\rangle = 0.
\end{equation}
Our data is then of the form:
\begin{equation}
	S = P_oW_o+X
\end{equation}
where
\begin{equation}
	X_{\mu i} \equiv \langle \vec f_\mu \cdot \vec g_i \rangle
\end{equation}
is the scalar noise correlation.

As a simple special case, consider $X = \chi \openone_8$ where $\chi \ll 1$.
We then have
\begin{equation}
S
\equiv
\left[
\begin{array}{cc}
	A & B\\
	C & D
\end{array}
\right]
+\chi\openone_8
=
\left[
\begin{array}{cc}
	A+\chi\openone_4 & B\\
	C & D+\chi\openone_4
\end{array}
\right]
\end{equation}
where $A^{-1} B D^{-1} C = \Delta(P_oW_o) = \openone_4$ so
\[
	\Delta^{-1} = \openone_4 + \chi(C^{-1} B^{-1} A + C^{-1} D B^{-1})
\]
and 
\begin{equation}
	\nabla = -\chi (\det D)\Big(A(BC)^{-1} A-\openone_4\Big).
\end{equation}
As expected, the entries of $\nabla$ are linear in the size of the correlations, $\chi$.

\subsubsection{Causal correlations}
A second model considers data of the form
\begin{equation}
	S_{\mu i} = \vec{p}_\mu\cdot\vec{w}_{i|\mu}
\end{equation}
where the states, $\vec{p}_\mu$, are constants but each detector has a value conditioned on the state they are detecting.
Correlations of this type are causal in the sense that the state preparation event is considered to be prior to the detection event and therefore the former cannot depend on the latter, but the latter can depend on the former.
More specifically, one could have, for example,
\begin{equation}
	\vec{w}_{i|\mu} = \vec{w}_{i o} + \chi\,\vec{p}_\mu
\end{equation}
where $\vec{w}_{i o}$ is fixed and $\chi \ll 1$ is a sort of linear susceptibility.
In this case, the data is of the form
\begin{equation}\label{susc}
	S = PW_o+Y
\end{equation}
where
\bea
	 P &\equiv&  
\left[
\begin{array}{ccc}
\cdots & \vec p_\mu & \cdots \\
	 \cdots& 1 &  \cdots
\end{array}
\right]^{T},
	 \nonumber\\
		 W_o &\equiv& 
\left[
\begin{array}{ccc}
\cdots & \vec w_{io} & \cdots \\
	 \cdots& 0 &  \cdots
\end{array}
\right]^{T},\\
	 Y_{\mu i} &\equiv& \chi |\vec{p}_\mu|^2.\nonumber
\eea
We can, obviously, construct additional more complex correlated models containing larger numbers of adjustable parameters. Fitting actual data to the different models and applying model selection would allow us to find the best description of the correlations present in our experiment (as explained further in Ref.~\cite{schwarz} in a similar context).
\section{Conclusions}
We considered a simple version of ``gate-set tomography'' \cite{gst}, in which only state preparations and measurements (SPAM) are performed, and no gates in between. As a special case we considered in detail qubits in $M$ {\em unknown} states measured by $N$ {\em unknown} two-outcome detectors.
These simplifications allowed us to fully analyze errors in SPAM and especially correlations between such errors. We found a simple test
for detecting such correlated errors in terms of a ``partial determinant'' (see Eq.~(\ref{pds}) or in terms of a nonsingular version, see Eq.~(\ref{nspd}), both of which are determined directly by the observed frequencies of the outcomes of the $M\times N$ different measurements. Such conditions
can be written down as soon as $M\geq 5$ and $N\geq 5$.

Finally, by considering the full set of gauge parameters (i.e., parameters not determinable by any (set of) SPAM experiment(s)), we found a novel interpretation of standard quantum state tomography and of the more recently developed detector topography: both can be seen as fixing the gauge parameters in their own particular way: the former by assuming some set of fiducial measurements is fully known, the latter by assuming some set of fiducial states is fully known.

\section*{Acknowledgments}
SJvE was supported in part by LPS/ARO under contract W911NF-14-C-0048.

\bibliography{spam_tomo}
\end{document}